\documentstyle[12pt,epsf]{article}

\begin{document}

\begin{titlepage}

\begin{flushright}
MIT-CTP-3205\\
hep-th/0111042
\end{flushright}

\vskip 2.5cm

\begin{center}
{\Large \bf Fermions in a Multi-Soliton Background\\
in 1+1 Dimensions}
\end{center}

\vspace{1ex}

\begin{center}
{\large B. Altschul\footnote{{\tt baltschu@mit.edu}}}

\vspace{5mm}
{\sl Department of Mathematics} \\
{\sl Massachusetts Institute of Technology} \\
{\sl Cambridge, MA, 02139-4307 USA} \\

\end{center}

\vspace{2.5ex}

\medskip

\centerline {\bf Abstract}

\bigskip

We consider the fermionic bound states associated with a soliton-antisoliton
pair in 1+1 dimensions which have zero energy when the solitons are infinitely
far apart. We calculate the energies of these states when the solitons are
separated by a finite distance. The energies decay exponentially with the
distance between the soliton and antisoliton.
When the fermion mass is much
larger than the boson mass, the energy simplifies substantially. 
These energies may be interpreted as a contribution to the effective potential
between the soliton and antisoliton. The character of this contribution depends
upon which fermionic states are occupied.
Performing the analogous calculation for the simplest (3+1)-dimensional soliton
system, we find no fermionic energy shift.

\bigskip 

\end{titlepage}

\newpage

The (1+1)-dimensional Dirac equation, when coupled to a scalar field, has the
interesting property that with the scalar field in a single-soliton state, the
fermionic spectrum can include a single zero-energy mode, so that the electric
charge of the system is half-integral~\cite{ref-jackiw}. However, while the
one-soliton case is exactly solvable, the problem of fermions in the presence
of multiple interacting solitons is difficult to analyze exactly. Graham
and Jaffe~\cite{ref-graham} have developed a numerical method for calculating
the fermionic energies in a multi-soliton background. We shall examine the
two-soliton case using an alternative method, deriving an analytic
approximation for the fermionic energies. Our approximation method can also be
applied to problems in 3+1 dimensions, where numerical calculations are
difficult.

The free Dirac equation in 1+1 dimensions is $i\gamma^{\mu}\partial_{\mu}\psi+
m\psi=0$, where a suitable choice for the Dirac matrices is $\alpha=\sigma_{2}$
and $\beta=\sigma_{1}$. If we use this representation, then the Hamiltonian $H=
\alpha p+\beta m$ becomes
\begin{equation}
H=-i\sigma_{2}\partial_{x}+\sigma_{1}m=\left[
\begin{array}{cc}
0 & -\partial_{x}+m \\
\partial_{x}+m & 0
\end{array}
\right].
\end{equation}
We shall replace the fermion mass $m$ with a scalar potential $U(x)$, which
is time-indepen\-dent (or varies very slowly in time), so that energy is
conserved.

We are interested in the case where the potential $U(x)$ that appears in
the Dirac equation is derived from a scalar field $\phi$. We shall take the
Lagrange density for this field to be
\begin{equation}
{\cal L}=\frac{1}{2}\dot{\phi}^{2}-\frac{1}{2}(\partial_{x}\phi)^{2}-V(\phi).
\end{equation}
We shall concentrate on the ``Mexican hat'' $\phi^{4}$ potential,
\begin{equation}
V(\phi)=\frac{\lambda^{2}}{4}\left(\phi^{2}-\frac{a^{2}}{\lambda^{2}}\right)
^{2},
\end{equation}
although similar results hold for for the sine-Gordon
potential~\cite{ref-rajaraman,ref-'thooft}.

The $\phi^{4}$ potential has two degenerate vacua, at $\phi=\pm\frac{a}
{\lambda}$. There are
also solutions that interpolate between the two vacua (i.e. solitons). For a
stationary soliton centered at $x=x_{0}$, the solution is
\begin{equation}
\label{eq-soliton}
\phi_{S,x_{0}}(x)=\frac{a}{\lambda}\tanh\left[\frac{\mu}{2}(x-x_{0})\right],
\end{equation}
where $\mu^{2}\equiv2a^{2}$ is the mass squared of the light
quanta of the scalar field. As $x\rightarrow\pm\infty$, $\phi_{S,x_{0}}(x)
\rightarrow\pm\frac{a}{\lambda}$. There is also an antisoliton solution
$\phi_{A,x_{0}}(x)=-\phi_{S,x_{0}}(x)$, which goes the other direction, from
$+\frac{a}{\lambda}$ at $x=-\infty$ to $-\frac{a}{\lambda}$ at $+\infty$.
Finally, there are solutions that resemble spatially
alternating solitons and antisolitons. However, treating these solutions as
collections of individual solitons and antisolitons is a good approximation
only when the solitons and antisolitons are widely separated, so that they
disturb one-another only minimally. In general, these solitons must be moving
relative to one-another, but we shall assume that this motion is very slow, so
that $\phi(x)$ depends very weakly on time. In three spatial dimensions, there
do exist exact multi-soliton (multi-monopole) static solutions to bosonic
field equations.

We may introduce massless fermions coupled to this scalar field by a
Yukawa interaction, $U(x)=g\phi(x)$. When $\phi$ is in a vacuum state, this
generates a fermion mass $m=g\frac{a}{\lambda}$. When $\phi$ has the soliton
profile $\phi_{S,x_{0}}$, there is known to be a zero-energy bound state
localized around the zero of the
soliton~\cite{ref-jackiw,ref-'thooft,ref-dashen}, with wavefunction
\begin{equation}
\label{eq-bound+}
\psi_{+}=C\left\{\cosh\left[\frac{\mu}{2}(x-x_{0})\right]\right\}^{-2m/\mu}
\left[
\begin{array}{c}
1 \\
0
\end{array}
\right].
\end{equation}
$C$ is a real normalization constant. For the antisoliton $\phi_{A,x_{0}}$ the
corresponding zero mode wavefunction is
\begin{equation}
\label{eq-bound-}
\psi_{-}=C\left\{\cosh\left[\frac{\mu}{2}(x-x_{0})\right]\right\}^{-2m/\mu}
\left[
\begin{array}{c}
0 \\
1
\end{array}
\right].
\end{equation}

We shall investigate how the energies of the fermion bound
states are affected in the two-soliton case. We shall consider an antisoliton
centered at $x_{1}$ and a soliton centered at $x_{2}$, where $\mu(x_{2}-x_{1})
\gg1$. We shall approximate $\phi$ as a sum $\phi=\phi_{A,x_{1}}+\phi_{S,x_{2}}
+\frac{a}{\lambda}$ and use $\psi_{+}$ and $\psi_{-}$ to represent the
bound state wavefunctions (\ref{eq-bound+}) and (\ref{eq-bound-}) centered at
$x_{2}$ and $x_{1}$, respectively.

When the solitons are far apart, we may expect the potential due to one soliton
to be a small perturbation of the Dirac equation for the bound state localized
around the other soliton. In
this spirit, we shall calculate the energy expectation values for the
symmetrized states $\Psi_{\pm}\equiv\frac{1}{\sqrt{2}}(\psi_{+}\pm\psi_{-})$.
For $H\psi_{+}$, we have
\begin{eqnarray}
H\psi_{+} & = & [-i\sigma_{2}\partial_{x}+\sigma_{1}g\phi_{S,x_{2}}]\psi_{+}+
\sigma_{1}g\left[\phi_{A,x_{1}}+\frac{a}{\lambda}\right]\psi_{+}
\nonumber\\
\label{eq-H+exact}
& = & 0 + \sigma_{1}[g\phi_{A,x_{1}}+m]\psi_{+}.
\end{eqnarray}

We shall first evaluate the energy expectation values $E_{\pm}\equiv\int
_{-\infty}^{\infty}dx\,\Psi_{\pm}^{\dag}H\Psi_{\pm}$ under the approximations
$\frac{\mu}{m}\gg1$ and $\mu(x_{2}-x_{1})\gg1$, where elementary expressions
can be obtained explicitly. Then we shall present an exact analysis for general
$\frac{\mu}{m}$, which reduces to the previous when $\frac{\mu}{m}\gg1$.

To find $E_{\pm}$, we
need to integrate quantities such as $(\psi_{+}^{\dag}\sigma_{1}\psi_{-})
\phi_{A,x_{1}}$. The characteristic size of $\phi_{A,x_{1}}$ is $\frac{1}
{\mu}$, and the characteristic decay length of $\psi_{\pm}$ is $\frac{1}
{m}$. If $\frac{\mu}{m}\gg1$, then the fermion wavefunctions decay very
little over the width of the antisoliton, and it is a good
approximation to replace $\phi_{A,x_{1}}$ by the step function $-\frac{a}
{\lambda}{\rm sgn}(x-x_{1})$ in the integral. Approximating $\phi_{A,x_{1}}$
with this step function gives
\begin{equation}
\label{eq-H+}
H\psi_{+}=\sigma_{1}m[1-{\rm sgn}(x-x_{1})]\psi_{+}.
\end{equation}
The corresponding expression for $H\psi_{-}$ is 
\begin{equation}
\label{eq-H-}
H\psi_{-}=\sigma_{1}m[1+{\rm sgn}(x-x_{2})]\psi_{-}.
\end{equation}

Combining (\ref{eq-H+}) and (\ref{eq-H-}), we get
\begin{equation}
H\Psi_{\pm}=\frac{1}{\sqrt{2}}\sigma_{1}m\{[1-{\rm sgn}(x-x_{1})]\psi_{+}\pm[1+
{\rm sgn}(x-x_{2})]\psi_{-}\}.
\end{equation}
Since $\psi_{+}^{\dag}\sigma_{1}\psi_{+}=\psi_{-}^{\dag}\sigma_{1}\psi_{-}=0$,
only the cross terms in $\Psi_{\pm}^{\dag}H\Psi_{\pm}$ are nonzero. Using the
relations
\begin{equation}
\psi_{+}^{\dag}\sigma_{1}\psi_{-}=\psi_{-}^{\dag}\sigma_{1}\psi_{+}=C^{2}
\left\{\cosh\left[\frac{\mu}{2}(x-x_{1})\right]\cosh\left[\frac{\mu}{2}(x-
x_{2})\right]\right\}^{-2m/\mu},
\end{equation}
the expression for $E_{\pm}$ is
\begin{eqnarray}
E_{\pm} & = & \pm m\frac{C^{2}}{2}\int_{-\infty}^{\infty}dx
\left\{\cosh\left[\frac{\mu}{2}\left(x-y+
\frac{\Delta x}{2}\right)\right]\cosh\left[\frac{\mu}{2}\left(x-y-\frac{\Delta
x}{2}\right)\right]\right\}^{-2m/\mu} \nonumber\\
& & \times\left[2-{\rm sgn}\left(x-y+\frac{\Delta x}{2}\right)+{\rm sgn}\left(x
-y-\frac{\Delta x}{2}\right)\right],
\end{eqnarray}
where $y\equiv\frac{x_{2}+x_{1}}{2}$, and $\Delta x\equiv x_{2}-x_{1}$ is the
separation between the solitons. Shifting he integration $x\rightarrow(x-y)$
gives
\begin{eqnarray}
E_{\pm} & = & \pm m\frac{C^{2}}{2}\int_{-\infty}^{\infty}dx
\left\{\cosh\left[\frac{\mu}{2}\left(x+
\frac{\Delta x}{2}\right)\right]\cosh\left[\frac{\mu}{2}\left(x-\frac{\Delta x}
{2}\right)\right]\right\}^{-2m/\mu} \nonumber\\
& & \times\left[2-{\rm sgn}\left(x+\frac{\Delta x}{2}\right)+{\rm sgn}\left(x-
\frac{\Delta x}{2}\right)\right],
\end{eqnarray}
which depends only on $\Delta x$. Thus we see that the zero-energy eigenstate
of the one-soliton background bifurcates into two states, symmetrically
displaced above and below $E=0$. To proceed further with our estimate of
$E_{\pm}$, we must approximate the wavefunctions, using
\begin{equation}
\label{eq-coshapprox}
\cosh\left[\frac{\mu}{2}(x-x_{0})\right]\approx\frac{1}{2}\left[\theta(x-x_{0})
e^{\frac{\mu}{2}(x-x_{0})}+\theta(x_{0}-x)e^{-\frac{\mu}{2}(x-x_{0})}\right].
\end{equation}
The approximation (\ref{eq-coshapprox}) is natural in this context, since the
approximate wavefunctions produced by (\ref{eq-coshapprox}) are the exact
wavefunctions corresponding to perfectly rectangular solitons. Using
(\ref{eq-coshapprox}), $E_{\pm}$ reduces to an integral over exponentials and
step functions, which may be evaluated exactly, yielding,
\begin{equation}
E_{\pm}\approx\pm(16)^{m/\mu}C^{2}e^{-m\Delta x}.
\end{equation}
It only remains to calculate $C$. Using the same approximation
(\ref{eq-coshapprox}) for cosh and demanding that $\int\psi_{+}^{\dag}\psi_{+}
\, dx=1$ gives
\begin{equation}
\label{eq-Capprox}
C^{2}\approx(16)^{-m/\mu}m,
\end{equation}
so that
\begin{equation}
\label{eq-solpot}
E_{\pm}\approx\pm me^{-m\Delta x}.
\end{equation}
An interaction energy of this nature has been previously
suggested~\cite{ref-jackiw}.

We are now in a position to show that the $\Psi_{\pm}$ are the correct
approximate states for which to calculate the energy. If we instead used some
other linear combination $(b\psi_{+}+c\psi_{-})$, with $|b|^{2}+|c|^{2}=1$, the
expectation value of the energy would be $E'=2({\rm Re}\{bc^{*}\})E_{+}$.
$E'$ takes its minimum value (corresponding to the best
approximation to the lowest-energy wavefunction) when $c=-b$. The best
approximation to the higher-energy wavefunction must be orthogonal to our
expression for the lower-lying state, and so has $c=b$. So the best linear
combinations of $\psi_{+}$ and $\psi_{-}$ are $\frac{1}{\sqrt{2}}(\psi_{+}\pm
\psi_{-})$.

To obtain the result (\ref{eq-solpot}), we needed to make the approximation
$\frac{\mu}{m}\gg1$. We now present an exact evaluation, which shows that the
above is the leading term for large $\frac{\mu}{m}$.

If we insert the exact expressions (\ref{eq-soliton}), (\ref{eq-bound+}), and
(\ref{eq-bound-}) into (\ref{eq-H+exact}) (and the corresponding expression for
$H\psi_{-}$), we get
\begin{eqnarray}
H\Psi_{\pm} & = & \frac{1}{\sqrt{2}}\sigma_{1}m\left\{1-\tanh\left[\frac{\mu}
{2}\left(x-y+\frac{\Delta x}{2}\right)\right]\right\}\psi_{+} \nonumber\\
& & \pm\frac{1}{\sqrt{2}}\sigma_{1}m\left\{1+\tanh\left[\frac{\mu}{2}\left(x-y-
\frac{\Delta x}{2}\right)\right]\right\}\psi_{-}.
\end{eqnarray}
Multiplying this by $\Psi_{\pm}^{\dag}$ from the left and integrating over
the shifted variable $(x-y)$ gives
\begin{eqnarray}
E_{\pm} & = & \pm m\frac{C^{2}}{2}\int_{-\infty}^{\infty}dx
\left\{{\rm sech}\left[\frac{\mu}{2}
\left(x+\frac{\Delta x}{2}\right)\right]{\rm sech}\left[\frac{\mu}{2}\left(x-
\frac{\Delta x}{2}\right)\right]\right\}^{2m/\mu} \nonumber\\
& & \times\left\{2-\tanh\left[\frac{\mu}{2}\left(x+\frac{\Delta x}{2}\right)
\right]+\tanh\left[\frac{\mu}{2}\left(x-\frac{\Delta x}{2}\right)\right]
\right\}.
\end{eqnarray}
When the integral is performed, the two tanh terms contribute equally, giving
\begin{eqnarray}
E_{\pm} & = & \pm mC^{2}\int_{-\infty}^{\infty}dx
\left\{{\rm sech}\left[\frac{\mu}{2}\left(x+\frac
{\Delta x}{2}\right)\right]{\rm sech}\left[\frac{\mu}{2}\left(x-\frac{\Delta x}
{2}\right)\right]\right\}^{2m/\mu}
\nonumber\\
\label{eq-Esech}
& & \times\left\{1+\tanh\left[\frac{\mu}{2}\left(x-\frac{\Delta x}{2}\right)
\right]\right\}.
\end{eqnarray}

We may simplify (\ref{eq-Esech}) using hyperbolic identities, in particular
\begin{equation}
\label{eq-sechid}
{\rm sech}(a+b)=\frac{{\rm sech}\,a\,{\rm sech}\,b}{1+\tanh a\tanh b}.
\end{equation}
Letting $a=\frac{\mu}{2}(x-\frac{\Delta x}{2})$ and $b=\frac{\mu}{2}\Delta x$
in (\ref{eq-sechid}) gives
\begin{equation}
\label{eq-Ealpha}
E_{\pm}=\pm mC^{2}\,{\rm sech}^{n}\left(\frac{m}{n}\Delta x\right)\int
_{-\infty}^{\infty}dx\,
\frac{{\rm sech}^{2n}\left[\frac{m}{n}\left(x-\frac{\Delta x}{2}\right)
\right]\left\{1+\tanh\left[\frac{m}{n}\left(x-\frac{\Delta x}{2}\right)
\right]\right\}}{\left\{1+\alpha\tanh\left[\frac{m}{n}\left(x-\frac{\Delta x}
{2}\right)\right]\right\}^{n}},
\end{equation}
where we have introduced the dimensionless parameters $n\equiv\frac{2m}{\mu}$
and $\alpha\equiv\tanh(\frac{m}{2}\Delta x)$. Making the substitution $u=\tanh
[\frac{m}{n}(x-\frac{\Delta x}{2})]$ in (\ref{eq-Ealpha}) simplifies this
integral substantially, giving
\begin{equation}
\label{eq-EwithC}
E_{\pm}=\pm mC^{2}\,{\rm sech}^{n}\left(\frac{m}{n}\Delta x\right)\int
_{-1}^{1}du\,\frac{n}{m}\frac{(1+u)^{n}(1-u)^{n-1}}{(1+\alpha u)^{n}}.
\end{equation}

To get a final expression for the energy, we must again calculate $C$. The
normalization gives
\begin{eqnarray}
\frac{1}{C^{2}} & = & \int_{-\infty}^{\infty}dx\, {\rm sech}^{2n}\left[\frac
{m}{n}(x-x_{0})\right] \nonumber\\
& = & \int_{-1}^{1}du\,\frac{n}{m}(1-u^{2})^{n-1} \nonumber\\
\label{eq-normint}
& = & \frac{n}{m}\sqrt{\pi}\frac{\Gamma(n)}{\Gamma(n+\frac{1}{2})},
\end{eqnarray}
which agrees with (\ref{eq-Capprox}) as $n\rightarrow0$.

The remaining integral in (\ref{eq-EwithC}) may also be found analytically. The
result involves the hypergeometric function  $F(a_{1},a_{2};b_{1};z)$. In terms
of this $F$, the integral is
\begin{equation}
\label{eq-hyperg}
\int_{-1}^{1}du\,\frac{(1+u)^{n}(1-u)^{n-1}}{(1+\alpha u)^{n}}=
\sqrt{\pi}\frac{\Gamma(n)}{\Gamma(n+\frac{1}{2})}(1+\alpha)^{-n}
F\left(n,n;1+2n;\frac{2\alpha}{1+\alpha}\right).
\end{equation}
Fortunately, this simplifies significantly when combined with
(\ref{eq-normint}), so that the final formula for the fermion energies is
\begin{equation}
\label{eq-Efinal}
E_{\pm}=\pm m\,{\rm sech}^{n}\left(\frac{m}{n}\Delta x\right)(1+\alpha)^{-n}
F\left(n,n;1+2n;\frac{2\alpha}{1+\alpha}\right).
\end{equation}

Since the solitons are far apart,
$\alpha\approx1-2e^{-\frac{2m}{n}\Delta x}\equiv1-\epsilon$ is close to one,
and it is natural to expand (\ref{eq-hyperg}) around $\alpha=1$. If we expand
the integral in (\ref{eq-hyperg}) to ${\cal O}(\epsilon)$, we find,
\begin{equation}
\label{eq-epsilon}
\int_{-1}^{1}du\,\frac{(1+u)^{n}(1-u)^{n-1}}{(1+\alpha u)^{n}}=
\int_{-1}^{1}du\,(1-u)^{n-1}+\epsilon\left[n\int_{-1}^{1}du\,\frac{u(1-u)
^{n-1}}{(1+u)}\right].
\end{equation}
The ${\cal O}(\epsilon^{0})$ term in (\ref{eq-epsilon}) has value $\frac{2^{n}}
{n}$. However, the ${\cal O}(\epsilon)$ term is divergent,
because $F(n,n;1+2n;z)$ is not analytic at $z=1$. Instead, we
have
\begin{equation}
\label{eq-epslog}
\int_{-1}^{1}du\,\frac{(1+u)^{n}(1-u)^{n-1}}{(1+\alpha u)^{n}}=
\frac{2^{n}}{n}+n2^{n-1}(\epsilon\ln\epsilon)+{\cal O}(\epsilon).
\end{equation}
Using (\ref{eq-epslog}) and $\epsilon=2e^{-\frac{2m}{n}\Delta x}$, the
${\cal O}(\epsilon\ln\epsilon)$ expression for $E_{\pm}$ is
\begin{equation}
E_{\pm}\approx\pm m\left[\frac{2^{n}}{\sqrt{\pi}}\frac{\Gamma(n+\frac{1}{2})}
{\Gamma(n+1)}\right]{\rm sech}^{n}\left(\frac{m}{n}\Delta x\right)\left[
1-n\left(2m\Delta x\right)e^{-\frac{2m}{n}\Delta x}\right].
\end{equation}
As $n\rightarrow0$, this agrees with the earlier result (\ref{eq-solpot}),
through terms of ${\cal O}(n)$.

The integral (\ref{eq-hyperg}) may be evaluated in terms of elementary
functions whenever $n$ is an integer. Of particular interest is the case $n=2$,
which corresponds to a supersymmetric Lagrangian. For that case,
the integral is
\begin{equation}
\label{eq-SUSY}
\int_{-1}^{1}du\, \frac{(1+u)^{2}(1-u)}{(1+\alpha u)^{2}}=\frac{1}{\alpha^{4}}
\left[(-4\alpha^{2}+6\alpha)+(\alpha^{2}+2\alpha-3)\ln\left(\frac{1+\alpha}
{1-\alpha}\right)\right],
\end{equation}
and the energy is
\begin{equation}
E_{\pm}=\pm\frac{3}{2}m\,{\rm sech}^{2}\left(\frac{m}{2}\Delta x\right)\left\{
\frac{1}{\alpha^{4}}\left[(-2\alpha^{2}+3\alpha)+\left(\frac{m}{2}\Delta x
\right)(\alpha^{2}+2\alpha-3)\right]\right\}.
\end{equation}
The first term on the right-hand-side of (\ref{eq-SUSY}) gives the ${\cal O}
(\epsilon^{0})$ term, while the second term is ${\cal O}(\epsilon\ln\epsilon)$.

The expression (\ref{eq-Efinal}) gives the energies for the fermion bound
states that lie closest to zero energy. When $n<1$, the
one-soliton system has only the single bound state 
(\ref{eq-bound+})~\cite{ref-dashen}, so the soliton-antisoliton system we are
considering has two bound states, with energies given by (\ref{eq-Efinal}). The
fermion energies depend on $\Delta x$, so (\ref{eq-Efinal}) generates a
contribution to the effective potential between the soliton and the 
antisoliton. The character of this potential term depends on which fermionic
states are occupied. In the vacuum, the state $\Psi_{-}$ is occupied, while
$\Psi_{+}$ is empty. $E_{-}$ is negative and for $\alpha\approx1$ becomes more
negative as $\Delta x$ decreases, creating
an attractive interaction between the soliton-antisoliton pair. Since
this interaction exists in the absence of any fermions or antifermions, it must
be generated by virtual particle effects. If either a
single fermion or a single antifermion is present (i.e. both states are either
filled or empty), the total energy vanishes, and these states do not
contribute to the effective
potential. Evidently, the presence of a single particle causes a cancellation
of the virtual particle corrections mentioned above. In the presence of both a
fermion and an antifermion ($\Psi_{+}$ occupied and $\Psi_{-}$ empty), the
potential term is repulsive. This presents an interesting picture, with the
fermions introducing novel interactions between the soliton-antisoliton pair.

If the positions of the soliton and antisoliton are reversed, so that $x_{1}>
x_{2}$, a calculation of the energies proceeds along the same lines. We
could extend (\ref{eq-Efinal}) by replacing $\Delta x$ by $|\Delta x|$
(both in the sech term and in $\alpha$) and multiplying by ${\rm sgn}(x_{2}
-x_{1})$. The energies obtained would be valid
whenever $\mu|\Delta x|\gg1$. However, there are additional subtleties that
arise when we allow $\Delta x$ to change signs. This involves the solitons
passing through one-another. The $\phi^{4}$ solitons do not pass through
one-another to emerge undistorted~\cite{ref-campbell}; a sine-Gordon soliton
and antisoliton will pass through each other without distortion, but when the
solitons are in precisely the same place, the field takes a vacuum value
everywhere, so there are no bound states at that instant~\cite{ref-rajaraman}.
In either case, it is impossible to identify which bound state for $x_{2}>
x_{1}$ corresponds to which bound state for $x_{1}>x_{2}$. However,
(\ref{eq-Efinal}) does have the attractive feature that the energies $E_{\pm}$
approach the energies of stationary continuum solutions as $\Delta
x\rightarrow0$ (even though (\ref{eq-Efinal}) is not valid for small values of
the separation).

The simplest (3+1)-dimensional soliton with a known zero-energy fermion mode
is the 't Hooft-Polyakov monopole~\cite{ref-jackiw,ref-'thooft-polyakov}
coupled to isospinor fermions (in the fundamental representation). The monopole
is described by a scalar field $\Phi_{M}=\Phi_{M}^{a}T^{a}$ and a vector field
$A_{M}^{0}=0$, $A_{M}^{i}=A_{M}^{ia}T^{a}$, where $T^{a}$ is the $2\times2$
isospin generator. Using the representation
\begin{equation}
\mbox{{\boldmath $\alpha$}}=\left[
\begin{array}{cc}
0 & \mbox{{\boldmath $\sigma$}} \\
\mbox{{\boldmath $\sigma$}} & 0
\end{array}
\right], \,\, \beta=i\left[
\begin{array}{cc}
0 & -I \\
I & 0
\end{array}
\right]
\end{equation}
for the Dirac matrices, the associated zero mode is
\begin{equation}
\psi_{M}=\left[
\begin{array}{c}
\chi_{+} \\
0
\end{array}
\right], \,\, \chi_{+}^{\nu n}=f(|{\bf r}-{\bf r}_{M}|)(s_{+}^{\nu}s_{-}^{n}-
s_{-}^{\nu}s_{+}^{n}),
\end{equation}
where $\nu$ is the spinor index, $n$ is the isospin index, ${\bf r}_{M}$ is the
position of the monopole, and $f$ is a known function of the boson fields. The
$s_{\pm}^{\nu,n}$ are orthonormal spinor, isospinor basis vectors. The
corresponding zero mode for an antisoliton with boson fields $\Phi_{A}$,
$A^{0}_{A}=0$, and $A^{i}_{A}$ is
\begin{equation}
\psi_{A}=\left[
\begin{array}{c}
0 \\
\chi_{-}
\end{array}
\right], \,\, \chi_{-}^{\nu n}=f(|{\bf r}-{\bf r}_{A}|)(s_{+}^{\nu}s_{-}^{n}-
s_{-}^{\nu}s_{+}^{n}).
\end{equation}

To apply our method to this system, we consider the approximate
monopole-antimono\-pole configuration $\Phi=\Phi_{M}+\Phi_{A}+\Phi_{0}$
(where $\Phi_{0}$ is the vacuum value of the field as $r\rightarrow\infty$),
$A^{\mu}=A^{\mu}_{M}+A^{\mu}_{A}$ and calculate the expectation value of the
energy for the state
\begin{equation}
\Psi_{\pm}\equiv\frac{1}{\sqrt{2}}(\psi_{M}+\psi_{A})=\frac{1}{\sqrt{2}}\left[
\begin{array}{c}
\chi_{+} \\
\chi_{-}
\end{array}
\right]
\end{equation}
in this background. This energy is
\begin{equation}
E_{\pm}=\frac{1}{2}\int d^{3}r\, [\chi^{\dag}_{+}\, \chi^{\dag}_{-}]\left[
\begin{array}{cc}
0 & \mbox{{\boldmath $\sigma$}}\!\cdot\!{\bf p}-\mbox{{\boldmath $\sigma$}}\!
\cdot\!{\bf A}^{a}T^{a}-iG\Phi^{a}T^{a} \\
\mbox{{\boldmath $\sigma$}}\!\cdot\!{\bf p}-\mbox{{\boldmath $\sigma$}}\!\cdot
\!{\bf A}^{a}T^{a}+ iG\Phi^{a}T^{a} & 0 \\
\end{array}
\right]\left[
\begin{array}{c}
\chi_{+} \\
\chi_{-}
\end{array}
\right],
\end{equation}
where $G$ is the strength of the Yukawa coupling. Since the states $\chi_{\pm}$
are singlets of spin plus isospin, the terms with  ${\bf p}$ and $\Phi$ vanish
immediately. The vector field term gives a contribution proportional to $\delta
_{i}^{a}A^{ia}$, which vanishes for the monopole-antimonopole profile, because
$A^{ia}$ is traceless in $(i,a)$, since it involves $\epsilon^{iaj}$. So in
this approximation, there are two degenerate zero-energy fermion modes for the
monopole-antimonopole system.

The isovector fermion does not obviously produce
zero-energy modes. Its properties are under investigation.

\section*{Acknowledgments}
This problem was mentioned by G. 't Hooft in reference~\cite{ref-'thooft}. The
author is also grateful to R. Jackiw and R. Jaffe for many helpful
discussions. This work is supported in part by funds provided by the U. S.
Department of Energy (D.O.E.) under cooperative research agreement
DE-FC02-94ER40818.

\end{document}